\documentclass[%
 reprint,showkeys,
nofootinbib,
 amsmath,amssymb,
 aps,
]{revtex4-2}
\usepackage{graphicx}
\usepackage{dcolumn}
\usepackage{bm}
\usepackage{hyperref}
\usepackage{derivative}

\usepackage{booktabs}
\newcommand\headercell[1]{%
   \smash[b]{\begin{tabular}[t]{@{}c@{}} #1 \end{tabular}}}
\begin{document}

\preprint{APS/123-QED}

\title{Estimating Stable Fixed Points and Langevin Potentials for Financial Dynamics}

\author{Tobias Wand}
\email{Corresponding author's e-mail address:\\ t\_wand01@uni-muenster.de}
\affiliation{Institut für Theoretische Physik, Westfälische Wilhelms-Universität Münster}
\affiliation{Center for Nonlinear Science Münster, Westfälische Wilhelms-Universität Münster}

\author{Timo Wiedemann}
\author{Jan Harren}
\affiliation{Finance Center Münster, Westfälische Wilhelms-Universität Münster}

\author{Oliver Kamps}
\affiliation{Center for Nonlinear Science Münster, Westfälische Wilhelms-Universität Münster}

\date{\today}

\begin{abstract}
The Geometric Brownian Motion (GBM) is a standard model in quantitative finance, but the potential function of its stochastic differential equation (SDE) cannot include stable nonzero prices. This article generalises the GBM to an SDE with polynomial drift of order $q$ and shows via model selection that $q=2$ is most frequently the optimal model to describe the data. Moreover, Markov chain Monte Carlo ensembles of the accompanying potential functions show a clear and pronounced potential well, indicating the existence of a stable price. 
\end{abstract}

\maketitle


\section{Introduction}

Research on financial data with methods from physics, summarised as \textit{Econophysics}, has lead to a better understanding of statistical properties such as financial correlation matrices \cite{MantegnaCorrelations,RandomMatrixLalouxPotters,Mnnix2012}, scaling behaviours of empirical distributions \cite{Mantegna1995}, microscopic trader models \cite{OrderSplitting_Farmer,Sato_Kanazawa} and other phenomena  \cite{MantegnaStanleyBook,Book_Econophysics_daCunha}. Differential equations such as the Brownian Motion and the Geometric Brownian Motion (GBM) in the Black-Scholes-Merton model have been an important tool to analyse financial data \cite{Bachelier,Black-Scholes,Merton}. Econophysics contributed to these efforts via ordinary (ODE), stochastic (SDE) and partial differential equations (PDE) \cite{RENNER2001499,Lorenz_Interest_Rates,Bouchaud_Cont_Langevin,Takayasu_FinancialBrownianMotion,Chapter_Estimate_GLE_Market_Price} and a recent empirical study modelled price time series with a harmonic oscillator ODE to reconcile the randomness of financial markets with the idea of a fair price \cite{GARCIA_physica_a_Harmonic_Oscillator_Finance}. \\
The GBM, still widely used as a standard model for price time series, presents the researchers with a subtle difficulty with regards to its interpretation: Its deterministic part implies either an unlimited exponential growth or an exponential decline to a price of 0 as pointed out in \cite{halperin2020inverted,Dixon2020,Halperin_QED}. While traditional finance models have tried to improve the GBM by changing its stochastic component, the deterministic part has largely been left unchanged (cf. the discussion in section 1 of \cite{Halperin_QED}). While \cite{Halperin_QED} used a constrained model with regularisation via strong prior information to fit parameters to their model, the goal of the present article is to estimate model parameters and to select the best model without any of these restrictions, i.e. letting the data speak for itself.\\
The estimation of Langevin equations from data via the Kramers-Moyal coefficients \cite{SIEGERT1998275,FRIEDRICH2000217} sparked a family of methods to estimate nonparametric drift and diffusion coefficients to model the observed system as a stochastic differential equation which have also been applied to financial data \cite{PhysRevLett.84.5224}. A particularly interesting expansion of this method is given by the maximum-likelihood-framework (ML) in \cite{KLEINHANS2007194}: for each time step $t_i$ and observed data $x_i$, the transition likelihood $L_i = p(x_{i+1} | x_i)$ from $x_i$ to $x_{i+1}$ is calculated and the joint likelihood $L = \sum_i L_i$ is maximised by the estimation algorithm. This approach takes the inherent stochasticity of stochastic differential equations into account and can be performed with a parametric model to recover algebraic equations increase its interpretability. Similarly, the SINDy algorithm recovers a sparse functional form of the underlying algebraic equations by fitting the data to a candidate function library, but struggles with noisy and stochastic data \cite{SINDy,Ensemble_SINDy}.\\

This article uses a combination of the ML framework of \cite{KLEINHANS2007194} with the candidate function library in \cite{SINDy} for a robust method to estimate stochastic differential equations from data similar to \cite{MartinHessler}. As an extension, the presented method can be used to estimate data from time series with non-constant time increments $dt_i \neq dt_j$. We use stock market prices at daily and 30-minute intervals as described in section \ref{sec:Data} to estimate their stochastic differential equations. In particular, we estimate the potential in which the dynamics take place to evaluate the stability of the dynamical process with the overall goal to distinguish between periods with a stable fixed point and unstable dynamics as explained in section \ref{sec:Theory}. The results for the different polynomial orders of the model and their implication for the stability are shown in section \ref{sec:Results} and discussed with respect to possible applications for risk assessment in section \ref{sec:Conclusion}.\\

\section{Data}
\label{sec:Data}
We analyse stock market data from the companies listed in table \ref{tab:tickers} to cover a range of different business sectors. Our analysis covers two distinct market conditions: (i) a calm period from early 2019 through early 2020 which was characterised by low overall volatility and (ii) the Covid Selloff  beginning March 2020 which was accompanied by a spike in market volatility. We analyse two sampling intervals: daily, end-of-day price changes (for which our data availability covers the whole of 2019 and 2020) and 30-minute intervals (for which our data is limited to the period between January 2019 up to and including July 2020).\\
Note that we are directly analysing the price time series $P_t$ instead of the returns $r = \log(P_{t+1}/P_t)$. Although analysis of the price data is also an important contribution to research \cite{COCHRANE2011}, the returns are often chosen as an observable because of their stationary distribution which allows the application of several time series analysis methods. However, our focus is explicitly on the non-stationary behaviour of stock prices: We estimate the potential of the differential equation's dynamics for different time intervals to differentiate between those dynamics with and without a stable fixed point (cf. section \ref{sec:Theory}). Similarly, the work in \cite{halperin2020inverted,Dixon2020,Halperin_QED} also uses prices to determine the position of the fixed points (or, equivalently, the wells of the potential): Although a return of $0$ also indicates a fixed point, it is not clear whether the price associated with it is the same as in the previous time window under observation. In particular, the research in \cite{halperin2020inverted,Dixon2020,Halperin_QED} stresses the important difference between fixed points at a nonzero price $P>0$ (normal behaviour of a stock) and at a price of $P=0$ (crash of the stock). Both phenomena correspond to a return of $r=0$, but describe vastly different situations of the stock.\\ 

\textbf{TAQ Database:} We use intraday data from the TAQ (Trade and Quote) database. To account for microstructure related issues, such as the bid-ask-bounce or infrequent trading, we rely upon quoted prices that we re-sample to a 30-minute frequency.\footnote{When we talk about quotes, we refer to the National Best Bid and Offer (NBBO) where the national best bid (offer) is the \textit{best} available quoted bid (offer) price across all U.S. exchanges. See \cite{holden2014liquidity} for an overview.} For that, we first remove all crossed quotes, i.e., all quotes where the bid price exceeds the ask, require the bid-ask-spread to be below 5\$, and finally use the last valid available quote within every 30-minute interval.\footnote{We forward fill quotes if there is no valid entry for a given time interval. However, this does almost never happen for very liquid stocks such as those chosen in this paper.} We further account for dividend payments and stock splits, which mechanically influence stock prices, and create a performance price index using quoted mid-prices. \\



 \textbf{CRSP:} We also consider lower-frequency, daily, data from the Center of Research in Security Prices (CRSP), which is one of the most used database in economics and finance. We again calculate a performance price index for each stock using the daily holding period return provided by CRSP. %
 Note that, while we use quoted mid-prices for the 30-minute high-frequency data, CRSP uses trade prices to calculate the holding period return. However, as the trading volume has increased considerably over the last decade, this should not be an issue \cite{wiedemann2022asynchronous}.

\begin{table}[]
    \centering
    \begin{tabular}{c|cc}
       Company  & Business Sector & Ticker \\ \hline
       Apple & Technology & AAPL \\
        Citigroup & Banking & C	\\
       Walt Disney Co. & Media & DIS \\
       Evergy Inc.& Energy & EVRG \\
       General Electrics & Industry &GE\\
       Pfizer& Pharmaceutics & PFE\\
       Walmart Inc. & Retail & WMT
    \end{tabular}
    \caption{The companies whose data has been analysed in our article.}
    \label{tab:tickers}
\end{table}

\section{Theoretical Background and Model}
\label{sec:Theory}
The standard stochastic differential equation to describe a stock price $P$ is the Geometric Brownian Motion given by
\begin{equation}
\label{eq:GBM}
 \odv{P}{t}   = \mu P + \sigma P \epsilon
\end{equation}
with standard Gaussian noise $\epsilon=\epsilon(t) \overset{iid}{\sim} \mathcal{N}(0,1)$, constant drift $\mu$ (typically $\mu>0$) and volatility $\sigma$. As pointed out in \cite{Halperin_QED}, the physical interpretation as a particle's trajectory $P$ in a potential $V(P)$ transforms equation \eqref{eq:GBM} to
\begin{equation}
\label{eq:GBM_Potential}
     \odv{P}{t}  =  - \odv{V}{P}(P) + \sigma P\epsilon_t \quad \textrm{with} \quad  V(P) = -\frac{\mu}{2}P^2
\end{equation}
and an arbitrary constant $C$ (set to zero for simplicity). However, analysing this potential $V$ in terms of its linear stability (cf. e.g. \cite{strogatz_book}) leads to the problematic result that the only fixed point in the data with $\odv{V}{P}(P_0) = 0$, namely  $P_0 = 0$, is an unstable fixed point for $\mu>0$. Without a stable fixed point, trajectories are expected to diverge away from $P_0 = 0$ towards infinity. As this is - at least for limited time scales - a highly unrealistic model, the authors of \cite{Halperin_QED,halperin2020inverted} have suggested higher order polynomials in the potential $V$ of \eqref{eq:GBM_Potential}. From the assumption that the rate of capital injection by investors should depend on the current market capitalisation, they derive a quartic potential
\begin{eqnarray}
\label{eq:GeneralPotential}
    V(P) & = &-P\left(\frac{\alpha_1}{2} P + \frac{\alpha_2}{3} P^2 + \frac{\alpha_3}{4} P^3\right)\textmd{ with } \nonumber \\
    - \odv{V}{P}(P) & = &\alpha_1 P + \alpha_2 P^2 + \alpha_3 P^3 ,
\end{eqnarray}
i.e. the drift term $- \odv{V}{P}(P)$ is a polynomial of order $q=3$. With a suitable choice of parameters $\alpha$, this potential can adopt the shape of a double-well potential with stable fixed points at $P_0=0$ and $P_1 > 0$, thereby predicting both the presence of a bankruptcy state at the stable fixed point $P_0=0$ and an additional stable state with nonzero price $P_1>0$. In \cite{Halperin_QED}, however, major constraints to the parameters during the estimation process were necessary to achieve this.

\subsection{Numerical Implementation}

If a time series $(s_n(t_n))_n$ with observations $s_n$ at time $t_n$ has been recorded, a maximum likelihood approach can be used to estimate the most likely parameters $\sigma$ and $\alpha_i$ such that a stochastic differential equation according to equations \eqref{eq:GBM_Potential} and \eqref{eq:GeneralPotential} may have produced the observed time series. For any two adjacent points $s_n(t_n)$ and $s_{n+1}(t_{n+1})$ and any given parameters $\phi = (\sigma^2, \alpha_i)$, the likelihood $L$ of observing the transition from $s_n(t_n)$ to $s_{n+1}$ at $t_{n+1}$ can be explicitly calculated as
\begin{eqnarray}
\label{eq:OneStepPropagator}
&L(s_{n+1}|t_{n+1}, s_n, t_n, \phi) =    \left( 2\pi  \left( \sigma s_n \sqrt{t_{n+1} - t_{n}}  \right)^2 \right)^{-\frac{1}{2}} \\ \nonumber
&\cdot\exp{ \left(  - \frac{\left( s_{n+1} -  \left(s_n + \left(-\odv{V}{s}(s) \right)(t_{n+1} - t_{n}) \right) \right)^2}{ 2 \left( \sigma s_n \sqrt{t_{n+1} - t_{n}}  \right)^2} \right)  } 
\end{eqnarray}
or as the log likelihood
\begin{eqnarray}
    &\mathcal{L}(s_{n+1}|t_{n+1}, s_n, t_n, \phi) = \log L(s_{n+1}|t_{n+1}, s_n, t_n, \phi)\\ \nonumber
    & = -\frac{1}{2} \log \left( 2\pi  \left( \sigma s_n \sqrt{t_{n+1} - t_{n}}  \right)^2 \right)\\ \nonumber 
    &- \frac{\left( s_{n+1} -  \left(s_n + \left(-\odv{V}{s}(s) \right)(t_{n+1} - t_{n}) \right) \right)^2}{ 2 \left( \sigma s_n \sqrt{t_{n+1} - t_{n}}  \right)^2}.
\end{eqnarray}
Because we assume Markovian dynamics, the complete log likelihood for the full observed time series is then simply the sum over the stepwise log likelihoods
\begin{equation}
\label{eq:LogLikelihood_full}
\mathcal{L}\left( (s_n)_n|(t_n)_n, \phi   \right)  = \sum_{i= 0}^{n-1}  \mathcal{L}(s_{i+1}|t_{i+1}, s_i, t_i, \phi).
\end{equation}
For given observations $(s_i, t_i)$, the likelihood $\mathcal{L}$ can be maximised by varying the parameters $\phi$ to estimate the optimal parameters $\phi^*$.\\

According to Bayes' Theorem and Bayesian Statsitics \cite{von2014bayesian}, the likelihood of observing the measured data conditional on some parameter values $L\left(\left(s_n(t_n)\right)_n\middle|\phi\right)$ is combined with an a-priori distribution $f_{prior}(\phi)$ to calculate a posterior distribution of the parameters given the observed data:
\begin{equation}
    \label{eq:Bayes}
    f_{post}(\phi | (s_n(t_n))_n) \sim f_{prior}(\phi) L\left(\left(s_n(t_n)\right)_n\middle|\phi\right).
\end{equation}
For an uninformed flat prior, this transformation is mathematically trivial, but allows us to calculate $f_{post}$ as a probability density of the parameters $\phi$ conditional on the observed data. Hence, the distribution of the parameters $\phi$ can be explored via Markov chain Monte Carlo (MCMC) methods (e.g. \cite{Hastings1970}) by drawing samples $(\phi^{(j)})_j$ from the posterior distribution as implemented in the Python package \textit{emcee} \cite{emcee}. MCMC can uncover correlations between different parameters and also explore local maxima of the probability density. It therefore gives a more complete view of the underlying distribution than summary statistics like e.g. the mean or standard deviation. In particular, we will use the sampled parameters $(\phi^{(j)})_j$ to construct an ensemble of potentials $V(P)$ and evaluate whether their shapes are roughly consistent with each other.

\subsection{Synthetic Data}
\label{sec:Synthetic}

To test our method, synthetic time series $(s_n)_n$ are simulated via the Euler-Maruyama scheme \cite{Kloeden} as

\begin{equation}
    \label{eq:Euler-Maruyama}
    s_{n+1} = s_n + \left(-\odv{V}{s}(s) \right)(t_{n+1} - t_{n}) + \sigma s_n \sqrt{t_{n+1} - t_{n}} \epsilon_n
\end{equation}
with $\epsilon_n \overset{iid}{\sim} \mathcal{N}(0,1)$ for any parameters $\alpha_i$ for the potential in \eqref{eq:GeneralPotential}. The following paragraphs discuss how well our model can then identify the underlying dynamcis and parameters from the observed data $(s_n,t_n)$. Note that for the synthetic data, non-equidistant time steps have been used.\\

\subsubsection{Estimating the Correct Order} 
\begin{figure}
    \centering
    \includegraphics[width = 0.45\textwidth]{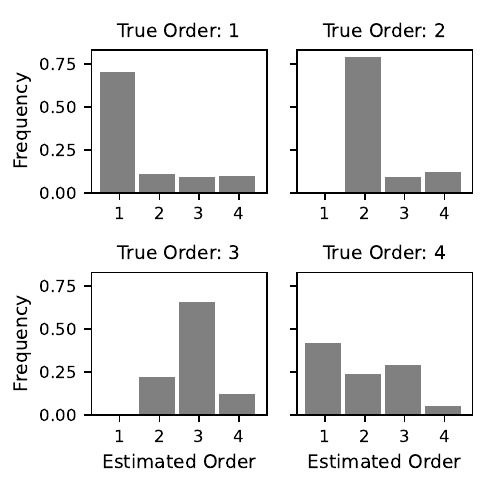}
    \caption{For each true polynomial order $q$, 100 trajectories are randomly sampled and then their best order is estimated. The histograms show that the method successfully estimates trajectories with order $q=1,2,3$, but struggles with the higher order $q=4$.}
    \label{fig:Histogram_of_Orders}
\end{figure}

Generalising the potential from \eqref{eq:GeneralPotential} to a potential with arbitrary polynomial order $q$ leads to
\begin{equation}
\label{eq:Potential_of_Order_q}
        V(P) = -P\sum_{i=1}^q\frac{\alpha_i}{i+1} P^i.
\end{equation}
For given $q$, random parameter values $\alpha_i$ and a random noise level $\sigma$ are sampled and the resulting time series is simulated with these parameters. If the sampled parameters result in numerical errors (i.e. an infinitely diverging time series), the time series is discarded from the ensemble. This is done repeatedly until the ensemble includes 100 time series with a length of 1000 time steps each. For those synthetic time series, the best order is estimated via the Akaike information criterion $AIC$ \cite{Akaike1998} given by 
\begin{equation}
    \label{eq:AIC}
    AIC  = -2\mathcal{L}_{\max} + 2(q+1)
\end{equation}
where $q+1$ is the total number of model parameters ($q$ monomials' prefactors $\alpha_i$ and $\sigma$). Hence, $q$ is varied, the maximum likelihood $\mathcal{L}_{\max}$ for the chosen $q$ is estimated and the resulting $AIC$ is calculated. The model with the lowest $AIC$ is chosen as the best model. The results shown in figure \ref{fig:Histogram_of_Orders} show that for polynomial orders $q=1$ (Geometric Brownian Motion), $q=2$ and $q=3$ (Halperin's suggestion as in equation \eqref{eq:GeneralPotential}), the correct order is usually identified as such. An even higher order $q=4$ shows very unreliable results, but will be included for completeness for the further data analysis.\\

\subsubsection{Estimating the Parameters} 
\begin{figure*}
    \centering
    \includegraphics[width = 0.9\textwidth]{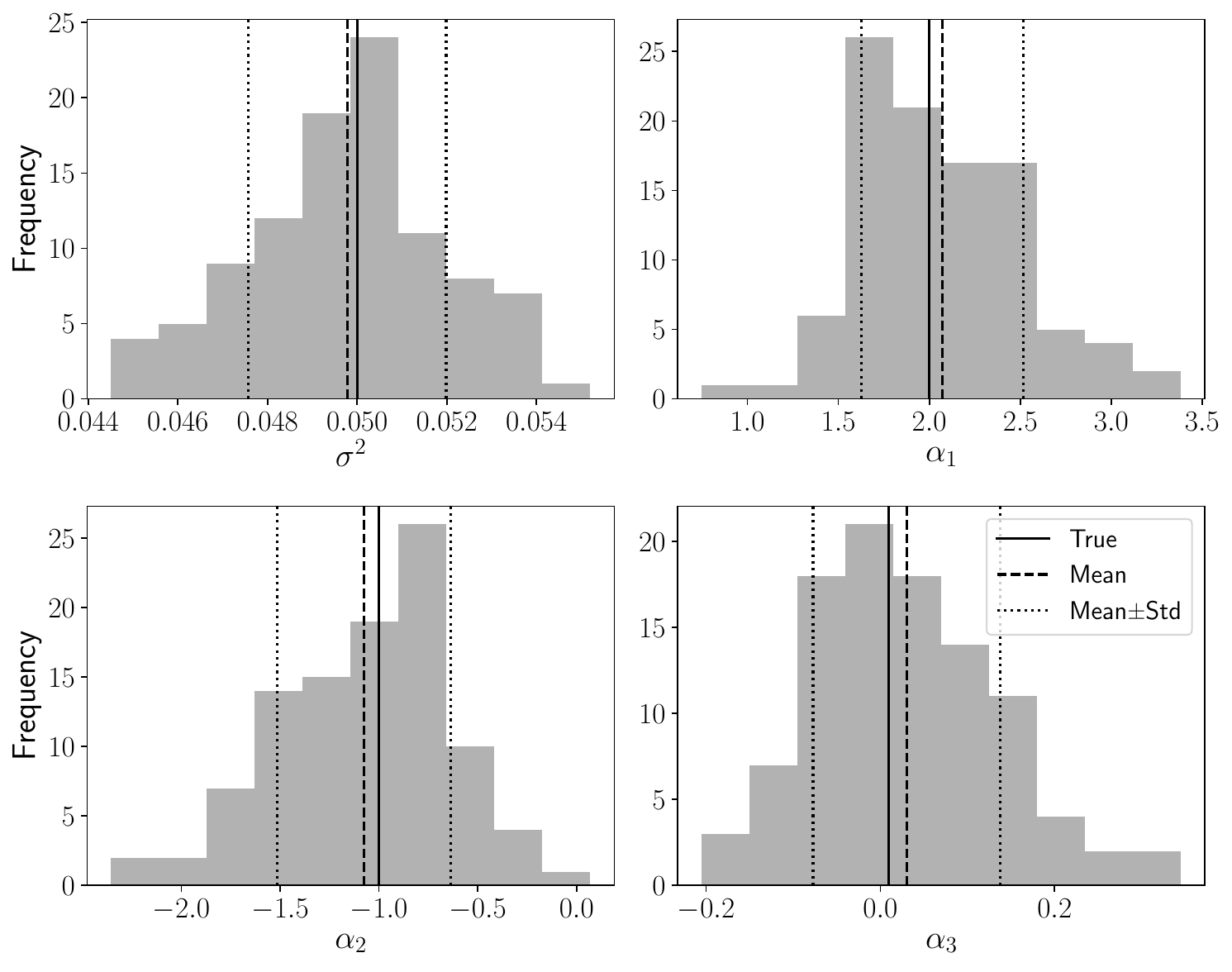}
    \caption{Parameter estimations for 100 trajectories with the same true parameters $\phi = (\sigma^2, \alpha_i)$ as given by the solid lines. The interval of Mean $\pm$ Standard Deviation of the estimated ensembles always includes the true parameters.}
    \label{fig:Parameters_for_order3}
\end{figure*}

Instead of sampling repeated trajectories with different parameters, now for order $q=3$, the parameters $\phi = (\sigma^2, \alpha_1, \alpha_2, \alpha_3)$ are kept constant as $\phi = (0.05, 2, -1, 0.01)$. 100 time series with 1000 time steps are sampled and their parameters are estimated by fitting a model with $q=3$. Despite the constant parameters, the randomness of the $\epsilon_n$ in equation \eqref{eq:Euler-Maruyama} nevertheless ensures that the time series are different from each other. The histograms of the estimated parameters, their means and standard deviations are shown together with the true parameter values in figure \ref{fig:Parameters_for_order3}. Note that the true parameter value is always within the one-standard-deviation-interval around the mean and that the parameter $\alpha_3$ has a distribution virtually indistinguishable from that of a parameter with mean zero: The parameter estimation correctly shows that $\alpha_3$ is so low that it is a superfluous parameter for model inference. Note that if the same data is estimated by a model with $q=2$, the results are fairly consistent with the depicted histograms. However, fitting the data to a model with $q=4$ results in the one-standard-deviation-interval of $\alpha_2$ also containing the value 0, which is a consequence of overfitting the model.

\section{Results}
\label{sec:Results}
 While \cite{Halperin_QED} analyses time periods of one year to estimate parameters, we believe that because of the assumption of constant volatility in equation \eqref{eq:OneStepPropagator}, it is prudent to restrict the date to shorter intervals of one trading month. Hence, we divide the given data into non-overlapping monthly intervals and estimate the polynomial order $q$ of the underlying stochastic differential equation via the $AIC$. Note that the time difference between each observation is taken as a constant interval of 1 time step in trading days or 30-minute-steps, respectively, including the overnight return.

\subsection{Polynomial Orders}
\begin{figure}
    \centering
    \includegraphics[width = \linewidth]{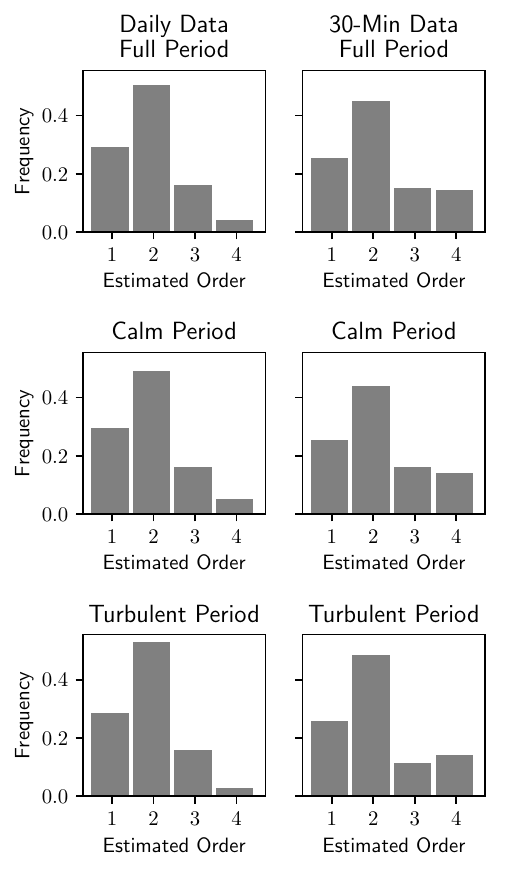}
    \caption{Estimated polynomial orders $q$ of the monthly real-world price time series of the companies in table \ref{tab:tickers} for the daily (left) and 30-minute-intervals (right). Differences between the calm (up to February 2020) and turbulent period (starting in March 2020) are only small. }
    \label{fig:Historgram_REALDATA_Orders}
\end{figure}

The distributions of the estimated polynomial orders $q$ are shown in figure \ref{fig:Historgram_REALDATA_Orders}: On both time scales and in all market periods, the order $q=2$ is the most frequently estimated order with the GBM model at $q=1$ being the second most frequent estimation. The suggestion $q=3$ from \cite{Halperin_QED} as well as the even higher-order $q=4$ are only rarely estimated as the most accurate model. Interestingly, calm and turbulent periods (as defined in section \ref{sec:Data}) show essentially identical distributions, whereas the order $q=4$ seems to be a bit more frequently estimated for the shorter time scale of 30 minutes than for the daily data. Table \ref{tab:ConfusionMatrix} shows that there is high consistency between the estimated orders for both time intervals for orders $q=1$ and $q=2$, but increasing disagreements for orders $q=3$ and $q=4$.\\ 
Overall, this suggests that a polynomial order of $q=1$ and $q=2$ can be a reasonable modelling assumption for the time series data and that the identification of these two orders is consistent for the two sampling intervals under consideration, whereas the choice of calm or turbulent periods does not seem to influence our results.\\

\begin{table}[]
    \centering
        \begin{tabular}{@{} c|cccc @{}}

\headercell{Optimal Order\\(30 Min)} & \multicolumn{4}{c@{}}{Opt. Order (Daily)}\\
\cmidrule(l){2-5}
 & 1 &  2 & 3 & 4   \\ 
\midrule
  1  & 24 & 6 & 3 & 1  \\
  2  & 6 & 41 & 10 & 3   \\  
  3  & 1 & 10 & 8 & 1   \\
  4  & 7 & 7 & 3 & 2  
\end{tabular}
    \caption{Comparison of the estimated orders $q$ for the same company and month with the price time series in daily and 30-minute-intervals.}
    \label{tab:ConfusionMatrix}
\end{table}

\subsection{Potentials}
For the optimal polynomial order $q$, we then sampled the parameters $\phi = (\sigma^2, \alpha_1,\dots, \alpha_q)$ from their posterior probability distribution to get an ensemble of parameters $(\phi^{(j)})_j$. From that, we calculate the corresponding ensemble of potentials $V(P)$ according to \eqref{eq:Potential_of_Order_q} and plot them, their pointwise centred 68\% and 95\% credible intervals (CIs) and the potential corresponding to the maximum likelihood estimation. The zero horizontal is shown in these plots as the y-axis position of the potential at $P=0$ to indicate where the potential is above or below the potential energy at the zero price (and if the price "particle" would therefore prefer or not prefer to be at the potential level of $P=0$). A couple of generic features can be observed for these potentials and do not depend on the chosen sampling rate:

\subsubsection{Order 2} 
\begin{figure}
    \centering
    \includegraphics[width=\linewidth]{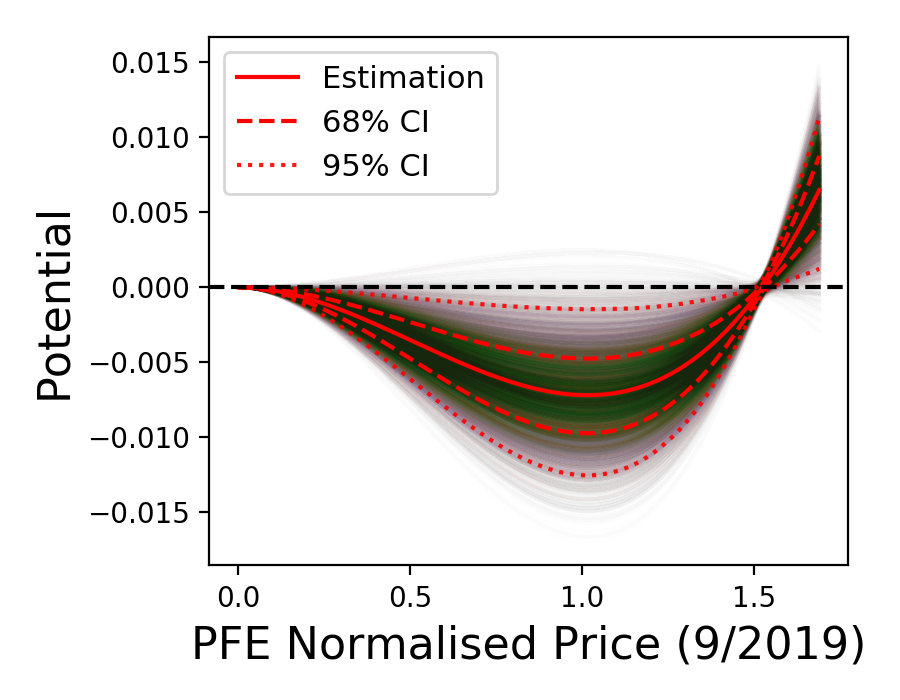}
    \caption{MCMC-sampled potentials of order $q=2$ for 30-minute-intervals for Pfizer in September 2019. The maximum likelihood estimation (Estimation) lies firmly within the ensemble of potentials and even the 68\% credible interval shows a clear potential well.}
    \label{fig:Potential_2nd_Order}
\end{figure}

As the order $q=2$ is the most frequently identified polynomial order according to the results in figure \ref{fig:Historgram_REALDATA_Orders}, it is quite insightful to focus on the associated potentials. They virtually always look like the potential depicted in figure \ref{fig:Potential_2nd_Order} and show a potential well as a pronounced minimum. Close to this minimum, the 68\% CI is usually also below 0 and sometimes (as depicted in figure \ref{fig:Potential_2nd_Order}) even the 95\% CI. The MCMC-sampled potential ensembles thus support the existence of a potential well as they clearly show the potential well for a large majority of trajectories. Following the interpretation of the potential wells from \cite{Halperin_QED}, this supports the existence of a locally stable price within this potential minimum. 

\subsubsection{Order 1: GBM}

\begin{figure*}
    \centering
\includegraphics[width=0.9\textwidth]{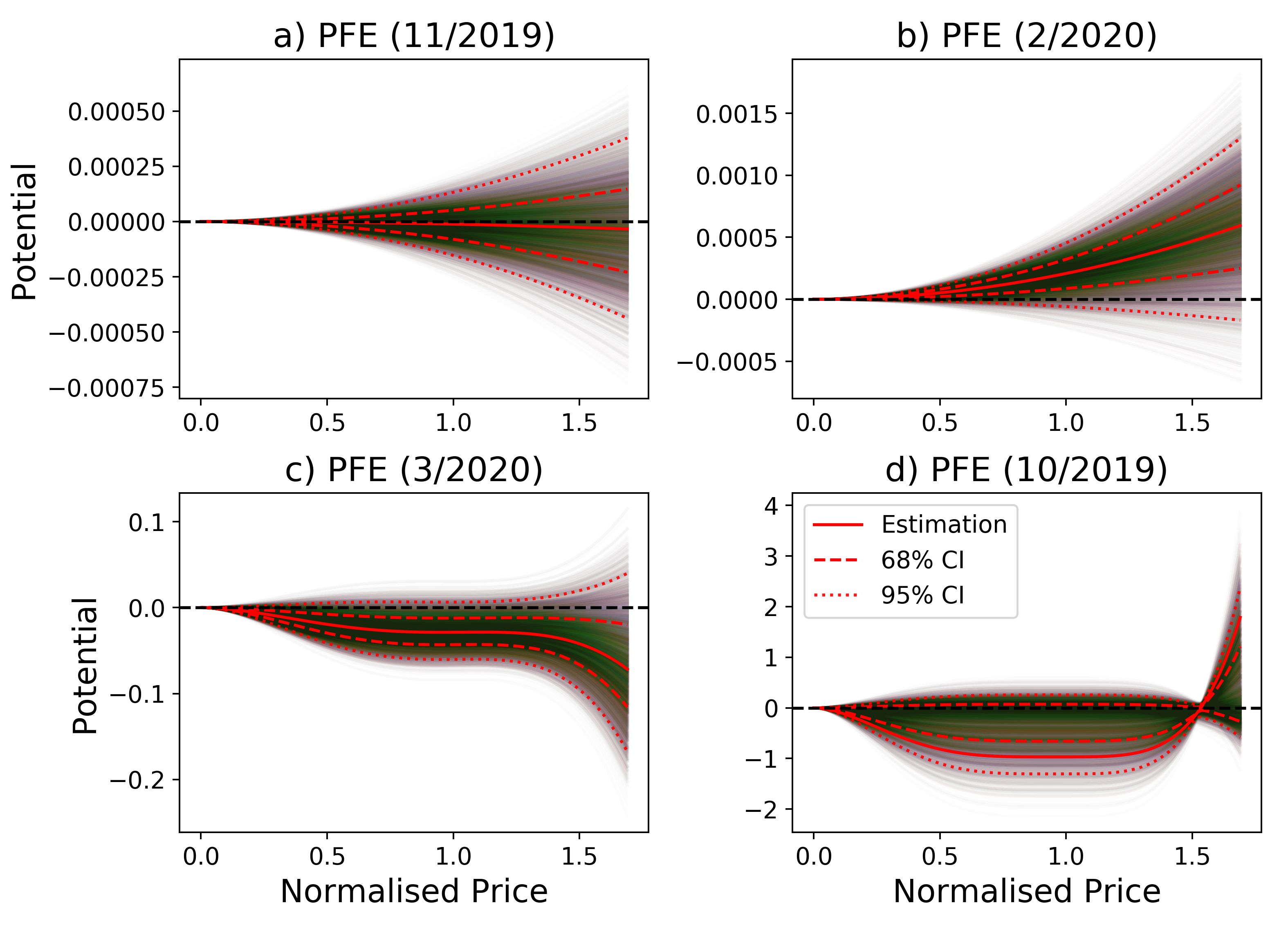}
    \caption{MCMC-sampled potentials of order $q=2$ for 30-minute-intervals for Pfizer in time intervals with $q=1$ ($a$ and $b$), $q=3$ ($c$) and $q=4$ ($d$). Note that the maximum likelihood estimation lies within the $68\%$ credible interval for all orders $q$ except for $q=4$.}
    \label{fig:ManyPotentials}
\end{figure*}

The GBM with $q=1$ is the second most frequently estimated  order. As shown in the subplots $a$ and $b$ in figure \ref{fig:ManyPotentials}, the MCMC samples show two general types of ensembles: in $a$, the maximum likelihood estimation of the potential is always very close to 0 and the 68\% CIs therefore envelope the zero horizontal. This makes it difficult to gauge a clear direction of the potential and hence of the movement of the price time series. Contrary to that, potential ensembles like in $b$ have a clear direction. In $b$, the potential is increasing for higher prices (and hence a restoring force pulls the price to the minimum at 0), but decreasing potential ensembles can also be found for other time intervals. That means some time intervals like $b$ show a predominant direction of price movement, whereas others like $a$ have no predominant direction, but rather a random movement.

\subsubsection{Order 3}
The potential with $q=3$ is the one suggested in \cite{Halperin_QED} and the typical shape of their MCMC samples are shown in subplot $c$ of figure \ref{fig:ManyPotentials}. Note that some of these potentials are also mirrored along the x-axis. Similar to the potentials in $b$, they also show a predominant direction, but also often a bistable saddle point. Notably, they do not show the pronounced double-well potential predicted in \cite{Halperin_QED}.

\subsubsection{Order 4}

Potentials with $q=4$ usually correspond to very wide potential wells as can be seen in subplot $d$ of figure \ref{fig:ManyPotentials} (compare e.g. its full width at half maximum to that of the potential in \ref{fig:Potential_2nd_Order}). In the depicted MCMC ensemble, the maximum likelihood estimation does not lie within the 68\% credible interval. This indicates a multimodal posterior distribution and was found in surprisingly many time intervals. Similar to the ensemble shown in subfigure $a$, these potentials also often envelope the x-axis with their 68\% CIs and therefore show no clearly predominant direction.

\section{Conclusion}
\label{sec:Conclusion}
\subsection{Summary}

We use a maximum likelihood estimation to analyse price time series of stocks. Via the $AIC$ model selection, we find that a second order polynomial for the drift term often offers a very suitable description of the data. While the standard GBM model with a first order polynomial is not selected as frequently as the second order model, it still appears often enough to be considered a valid candidate model. Higher order polynomials are rarely estimated. Sampling the posterior density of the parameters via MCMC reveals that the second order polynomials' potentials show very pronounced potential wells (i.e. stable minima) for nonzero prices which is mathematically impossible for the GBM's potential as pointed out in \cite{Halperin_QED}.

\subsection{Discussion}

\begin{figure}
    \centering
    \includegraphics[width=\linewidth]{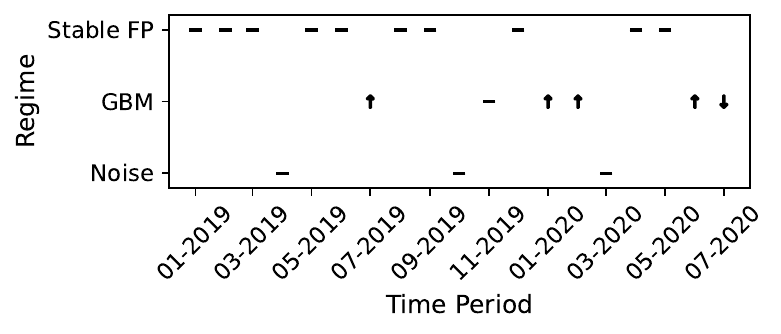}
    \caption{Evaluation of the different regimes for the dynamics of the Pfizer stock for 30-minute-intervals. the rare orders $q=3$ and $q=4$ have been summarised under the label ''Noise'', the label ''Stable FP'' indicates order $q=2$ with a potential well and the GBM of order $q=1$ is further split up into periods of growth ($\uparrow$), random stagnation (--) and decline ($\downarrow$): according to their MCMC-sampled potentials (cf. subfigures a and b in \ref{fig:ManyPotentials}), a growth or decline is only assumed if the $68\%$ CIs do not include the 0 horizontal.}
    \label{fig:PFE_30_Min_Regimes}
\end{figure}

Our research question is heavily inspired by \cite{Halperin_QED}, but differs from it in a key factor: The model presented in \cite{Halperin_QED} always has a drift polynomial of order $q=3$ and uses the credit default swap rates (CDS) to estimate the probability of a considered company going bankrupt. This probability is then used as a constraint in the parameter estimation such that the jump from a potential well with nonzero price to another potential well at price zero (i.e. the stock collapsing) has a jump probability (Kramers's escape rate) which is quantified by the CDS. Thus, the work in \cite{Halperin_QED} combines the price dynamics of stochastic differential equations with the CDS data as additional constraints to estimate a stochastic differential equation with a probability of the stock crashing. In contrast to this, our estimation scheme uses no additional constraints or external data, but purely the price time series. Our $q=3$ estimations (subfigure $c$ in \ref{fig:ManyPotentials}) do not show the double-well potential postulated by \cite{Halperin_QED}. However, our model selection via the $AIC$ indicates that $q=2$ is instead the most frequently observed polynomial degree and for $q=2$, MCMC shows a clear potential well that is consistent for the whole MCMC ensemble. In short, because we do not use additional constraints, we cannot reproduce the double well potential with default probabilities, but instead show via our fully unconstrained approach that the potential wells arise naturally just from the price time series alone. Estimating potentials for stochastic financial dynamics and analysing the stability of their fixed points has also been done in \cite{Rinn2015DynamicsOQ,Stepanov2015}, but two key differences exist between them and our approach: While we estimate explicitly analytical potentials for the time series of individual stocks, the work in \cite{Rinn2015DynamicsOQ,Stepanov2015} estimates potentials purely numerically without a closed-form analytical expression and does so for the collective market movement instead of treating individual assets. Interestingly, \cite{Rinn2015DynamicsOQ,Stepanov2015} observe transitions between the different minima of the potentials and therefore a non-stationary market behaviour, similar our study, because the $AIC$ selection means that the stocks are not described by the same polynomial degree $q$ for all time series. Instead, our model selection implies that the potential itself is time-dependent. \\
A possible explanation for this is that due to external effects, an order parameter such as the capital influx into the financial markets is changed. Then, the underlying potential might change due to these effects and e.g. experience a bifurcation, resulting in changing price dynamics. Whether a bifurcation is a suitable description of the dynamics under such conditions requires further analysis on the transitions between the different models. One can imagine e.g. that a price initially starts as being in a stable fixed point with $q=2$ like in figure \ref{fig:Potential_2nd_Order}, but external news change the potential to that of subfigure $b$ in figure \ref{fig:ManyPotentials} with $q=1$. Now, the price has a predominant direction of movement and is no longer experiencing a restoring force back to the price at the previous fixed point and can therefore explore new areas of the phase space (figure \ref{fig:PFE_30_Min_Regimes} illustrates the transition between different regimes). The market finally manages to process the news and their implications and finally, the price reaches a new fixed point with $q=2$. Thus, the price at the potential minimum can be interpreted as a fair price similar to the discussion in \cite{GARCIA_physica_a_Harmonic_Oscillator_Finance}. However, further research into the transition between the different potentials is necessary to verify this interpretation. Note that GBMs with potentials such as subfigure $a$ in figure \ref{fig:ManyPotentials} have essentially no predominant direction of movement and show a random walk without a restoring force. This is a different behaviour to $q=2$ which also does not show a predominant direction, but instead has such a restoring force that restricts the price to the potential well.    \\
One might have assumed that stable fixed points ($q=2$) should occur significantly less frequently during the turbulent period because of the overall instability of the market. But interestingly, our results do not seem to show a difference between the calm and turbulent market period (cf. figure \ref{fig:Historgram_REALDATA_Orders}), perhaps indicating that the market can quickly adjust to such turbulent behaviour.\\ Finally, it is a reassuring result that the $AIC$ selection still frequently suggests $q=1$ (the standard GBM model) as the best polynomial order. The standard GBM model still appears rather frequently in our data and therefore nevertheless manages to provide a reasonably accurate model.

\subsection{Further Research and Applications}
As discussed in the previous subsection, our method can be used to distinguish between different regimes (stable fixed point or growth/decay) of the dynamics of a stock time series. One could use our methodology to continuously model a given time series, update it with new data and pay attention to when the potential is changing such that the system is transitioning from a stable (resilient) state to an unstable one or vice versa. This point of view can be used to judge the system's resilience against noise and anticipate critical transitions to a qualitatively new system behaviour \cite{MartinHessler,Hessler_Nexus}. In the non-stationary system of a free market, such monitoring might support risk management decisions.\\ 

While this article focused on the drift term like in \cite{Halperin_QED}, there are of course possible extensions of the diffusion/volatility that can be taken into account, too. Stochastic volatility and local volatility models have been widely accepted in finance \cite{StochasticVola_heston1993closed,LocalVola_dupire1994pricing}, but other modelling possibilities exist, too: While our article used the volatility parametrisation from the GBM in \eqref{eq:GBM} via $\sigma P \epsilon_t$, one can also imagine e.g. a polynomial model here given by 
\begin{equation}
    \textmd{Diffusion}(P) = \sigma \left( \sum_j \beta_j P^j \right) \epsilon_t.
\end{equation}
However, from the author's experience, the maximum likelihood estimation can become troublesome if the diffusion term has several free parameters as the estimator can then attempt to essentially attribute the whole observed dynamics to the diffusion. A strong regularisation might be necessary if one wishes to expand the diffusion model. A multi-stage estimation procedure might provide another alternative: First estimate the GBM model with drift parameters $\phi_{D,1}$, then keep these parameters fixed to estimate the parameters $\phi_{V,1}$ of a more complicated volatility model (e.g. Heston's stochastic volatility). Then keep the parameters $\phi_{V,1}$ of the volatility model fixed and vary the drift parameters according to the scheme presented in our article in order to find the optimal order $q$ and its associated parameters $\phi_{D,2}$. For fixed order $q$, iteratively use fixed $\phi_{D,n}$ to estimate $\phi_{V,n+1}$ and fixed $\phi_{V,n+1}$ to estimate $\phi_{D,n+1}$ until the parameter values converge. Developing and fine-tuning this procedure, however, is beyond the scope of the present work whose main aim was to investigate the existence of stable fixed points in the drift potential.\\
Another model extension might be the incorporation of memory effects. Generalised versions of the Langevin equation include non-Markovian memory terms by e.g. an explicit memory kernel \cite{GLE_Clemens} or by assuming the existence of a second hidden process that has not been observed \cite{Willers2021}. Such a hidden component might correspond to the traders' knowledge or belief which certainly influences the stock prices, but is not explicitly recorded. Although we believe that there is some virtue in having a simple model as evidenced by the widespread use of the GBM, a more complex analytical model than a polynomial approach can of course be used in the maximum likelihood framework to expand our rather simple model. Combining all those extensions and using a strict regularisation procedure to discard superfluous terms might ultimately help to develop a model that not only differentiates between the different regimes of stability (as shown in the present article), but also reproduces the well-known stylised facts from the empirical literature. \\
Finally, it is noteworthy to point out that although we used equidistant time intervals between the observations of the data, the model has been tested on synthetic data with non-equidistant time intervals in section \ref{sec:Synthetic}. Such a situation arises naturally in the context of tick-by-tick data which is the highest resolution of trading data. Here, instead of sampling the price at a high frequency, every single trade is recorded at the exact time that it occurred. As the time between two subsequent trades can be arbitrarily short or long, the application of a robust method without the need for equidistant time steps might prove useful here.

\begin{acknowledgements}
    The authors thank Tim Kroll (WWU Münster) for valuable discussion about the propagator and its technical implementation and the anonymous reviewers for their valuable advice. Tobias Wand is financed by the Studienstiftung des deutschen Volkes.
\end{acknowledgements}

\bibliography{apssamp}

\end{document}